\documentclass[twoside]{article}
\usepackage{fleqn,espcrc2,epsfig}

\title{Recent Results From the K2K (KEK-to-Kamioka) Neutrino Oscillation Experiment}

\author{S. Boyd\address[UW]{University of Washington,  Seattle}}

\begin{document}

\begin{abstract}
The latest results of the K2K experiment are reported. The results
are based on data taken from June, 1999, to June, 2000, corresponding to
a total $2.6 \cdot 10^{19}$ protons on target.
Twenty seven fully-contained events in the
22.5 kton fiducial volume of Super-Kamiokande (SK) are observed. The
expected number of events is estimated 
to be $40.3^{+4.7}_{-4.6}$ assuming the null oscillation hypothesis.
\vspace{1pc}
\end{abstract}

% typeset front matter (including abstract)
\maketitle

\section{Introduction}
\label{sec:intro}

 For many years experiments studying neutrinos produced in interactions
of cosmic rays with atmospheric nuclei have been measuring a significant
suppression of the ratio of $\nu_{\mu}$ to $\nu_{e}$ with respect to the expectation derived from
standard hadronic shower models \cite{kam1,kam2,sk1,sk2,imb,soudan}. The popular interpretation
of this suppression is that neutrino oscillations, in which $\nu_{\mu}$ change
to some other
neutrino flavour in flight, are occurring. Neutrino oscillations can only
take place if the neutrino has a finite mass and if, as in the quark sector,
the neutrino mass eigenstates are not the same the flavour eigenstates.

The probabilty of a neutrino with energy $E_{\nu}$(GeV) produced a distance
$L$(km) from a detector to oscillate into another flavour is expressed (in the two-flavour
approximation) as :
\begin{equation}
 P(\nu_{\mu}\rightarrow\nu_x)=sin^22\theta sin^2
\frac{1.27\Delta m^2 (eV^2)L(km)}{E_{\nu}(GeV)}
\end{equation}
where $\Delta m^2$ is the difference between the squared masses of the two neutrino
mass eigenstates and $sin^22\theta$ is a mixing parameter. 
The recent results
from the Super-Kamiokande experiment \cite{SKatm:ref} have presented strong
evidence that the atmospheric neutrino problem can be described assuming
the existence of neutrino oscillations with $\Delta m^2$ in the range $2 \cdot 10^{-3}$ to
$6 \cdot 10^{-3}$ and $sin^22\theta > 0.8$ at the 90\% confidence level.

The K2K experiment is the first accelerator-based long baseline neutrino oscillation
experiment. With a baseline of 250 km, an average neutrino energy of E$\sim$1.3 GeV
and an intense neutrino beam which is 98.0\% pure $\nu_{\mu}$, K2K is ideally placed
to investigate neutrino oscillations in the oscillation parameter region favoured
by Super-Kamiokande. The primary oscillation search mode is $\nu_{\mu} \rightarrow \nu_{\tau}$
oscillations (the $\nu_{\mu}$ disappearance mode). 
However K2K can also investigate flavour oscillations through the $\nu_{\mu} \rightarrow \nu_{e}$
appearance mode.

In this paper, the most recent results from K2K are presented. Section \ref{sec:k2k} describes
the beam and the near detector components. Section \ref{sec:fgd} discusses the measurement
technique and the calculation of the expected flux of neutrinos at Super-Kamiokande. Section
\ref{sec:sk} outlines the selection of KEK beam-related events in Super-Kamiokande, compares
the number of events observed in Super-Kamiokande with the expected number and describes the
systematic errors in the measurement. Finally, a conclusion is drawn in Section \ref{sec:concl}.

\section{K2K Experiment}
\label{sec:k2k}
\subsection{Beamline and beam monitors}

\begin{table}
\label{table:beamcomp}
\caption{Flavour composition of the KEK wide-band neutrino beam.}
\begin{tabular}{lc}
\hline\noalign{\smallskip}
  Neutrino Flavour   &   Flux Abundance \\
\noalign{\smallskip}\hline\noalign{\smallskip}
$\nu_{\mu}$            &  98.0\% \\
$\overline{\nu}_{\mu}$ &  1.0\% \\
$\nu_{e}$            &  1.0\% \\
\noalign{\smallskip}\hline
\end{tabular}
\end{table}

The KEK neutrino beam is produced by 12 GeV/c protons incident on an aluminium target.
The 66 cm long aluminium target is embedded in the first of two focussing horns which serve to
focus(defocus) the positive(negative) pions and kaons produced by the proton-Aluminium interactions.
The positive hadrons then passed to a 200~m long decay pipe where
they decay, producing a wide-band neutrino beam with a mean energy of 1.3 GeV. The beam composition
estimated from Monte Carlo simulation is shown in Table \ref{table:beamcomp}. 
The beam Monte Carlo program itself uses the proton beam profile before
the target and generates $\pi^{\pm,0} $, $K^{\pm,0}$ and secondary protons
according to a parametrisation of existing data on secondary particle production in p-Al interactions\cite{Cho}.
GEANT 3.15\cite{GEANT:ref} is then used to trace the secondary particles through the target,
the magnetic horns, and into the decay tunnel.

\begin{figure}
\centerline{\psfig{file=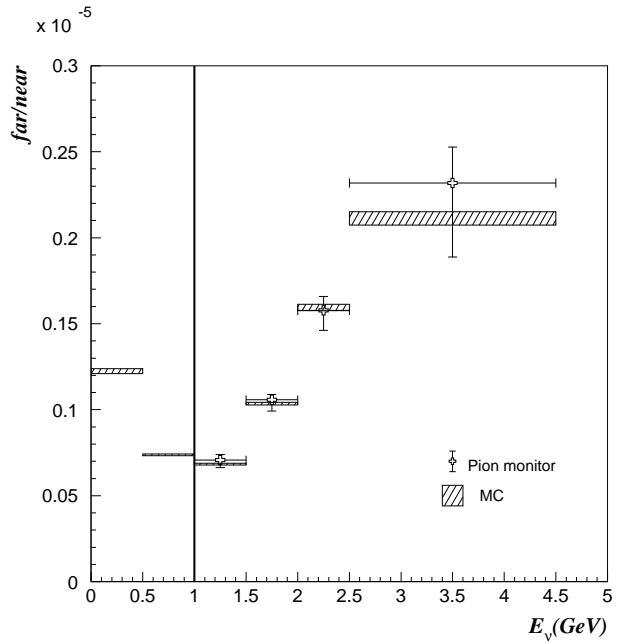,width=8cm}}
\caption{Far to near flux ratio predicted by the pion monitor compared with the prediction from 
the beam simulation.
The pion monitor is sensitive only above a neutrino energy of 1 GeV.}
\label{fig:pion}
\end{figure}

 Within the target hall, and before the decay pipe, a threshold gas Cerenkov detector (the pion monitor)
has been constructed. This monitor can be moved in and out
of the beam and is used to measure the energy and angular spectra
of the secondary pions after focussing. The measurements are then used to verify the predictions
of the beam Monte Carlo. Figure \ref{fig:pion} compares the ratio of the $\nu_{\mu}$ flux seen
in the far detector (see Section \ref{sec:fgd}) and in the near detector as predicted by the pion monitor
and by the beam simulation. Above a neutrino energy of 1 GeV (the lower limit of the pion monitor sensitivity)
the pion monitor predictions and the beam simulation agree.

At the end of the decay pipe is an iron and concrete beam dump after which is positioned a
segmented plate ionisation chamber and a set of silicon detectors. These monitor the
residual beam-associated muons and are used to check the beam centering and intensity.

The direction from KEK to Super-Kamiokande along which to point the beam was surveyed using
a method based on GPS\cite{noumi}. The survey precision is $<\rm 0.01\,mrad$ and the
construction precision is $<\rm 0.1\,mrad$. Due to beam divergence over the 250 km flight
path, the flux of neutrinos at Super-Kamiokande is constant over a distance of approximately one km
around Super-Kamiokande, requiring a pointed accuracy of $< 3\,mrad$, well above the estimated
accuracy.

\subsection{The near detector}

 In a disappearance oscillation experiment it is essential that the neutrino beam at the
point of production be understood completely. The understanding of the neutrino beam 
is the
primary goal of the suite of detectors located 300~m downstream of the target. The near
detector is composed of a one kiloton water Cerenkov detector (1kt) and a fine-grained
detector (FGD).

 The 1kt detector measures neutrino interactions using the
same target and techniques as used in the far detector (Super-Kamiokande) thus cancelling many of
the systematic errors in the calculation of the expected event rate in the far detector
(see Section \ref{sec:fgd}).

 The FGD complements the one kiloton detector. It is used for precise
measurements of the $\nu_{\mu}$ energy spectrum and flux, the beam profile and the
$\nu_{e}$ contamination in the $\nu_{\mu}$ beam. The fine-grained detector consists of 
a scintillating fiber
tracking detector (SciFi)\cite{scifiNIM}, a lead-glass electromagnetic calorimeter and a
large muon range detector (MRD).

 The SciFi is constructed from 20 $2.4m \times 2.4m$ modules. Between each module is a
6~cm thick layer of aluminium tubes filled with water. These form the target for neutrino
interactions in the SciFi. Such interactions are used to study the neutrino energy spectrum and for
studying other properties of neutrino-water interactions at the 1 GeV energy scale.

 Downstream of the SciFi is the lead-glass array. The purpose of the lead-glass array is to
identify $\nu_{e}$ interactions. The energy resolution of the lead-glass detector is
$\Delta E/E=10\%/\sqrt{E(GeV)}$. 

 The muon range detector (MRD) is a large ($7.6~m \times 7.6~m$ cross-sectional area) iron tracking
calorimeter with a target mass of 915 tons. 
It consists of 12 iron plates interleaved with drift chambers. The four most upstream
plates are 10~cm thick and the last eight plates are 20~cm thick. The MRD is used to
monitor the beam profile, to study the stability of beam intensity  and to 
determine the energy of muons from neutrino interactions
in the MRD or SciFi using rangeout techniques. 

The far detector for K2K is the Super--Kamiokande detector\cite{SKatm:ref}
which has been taking data since 1996. The data selection at this detector is described
below in Section \ref{sec:sk}.

\section{Performance of the near detector and beam stability}
\label{sec:fgd}

\begin{figure}[p]
\centerline{\psfig{file=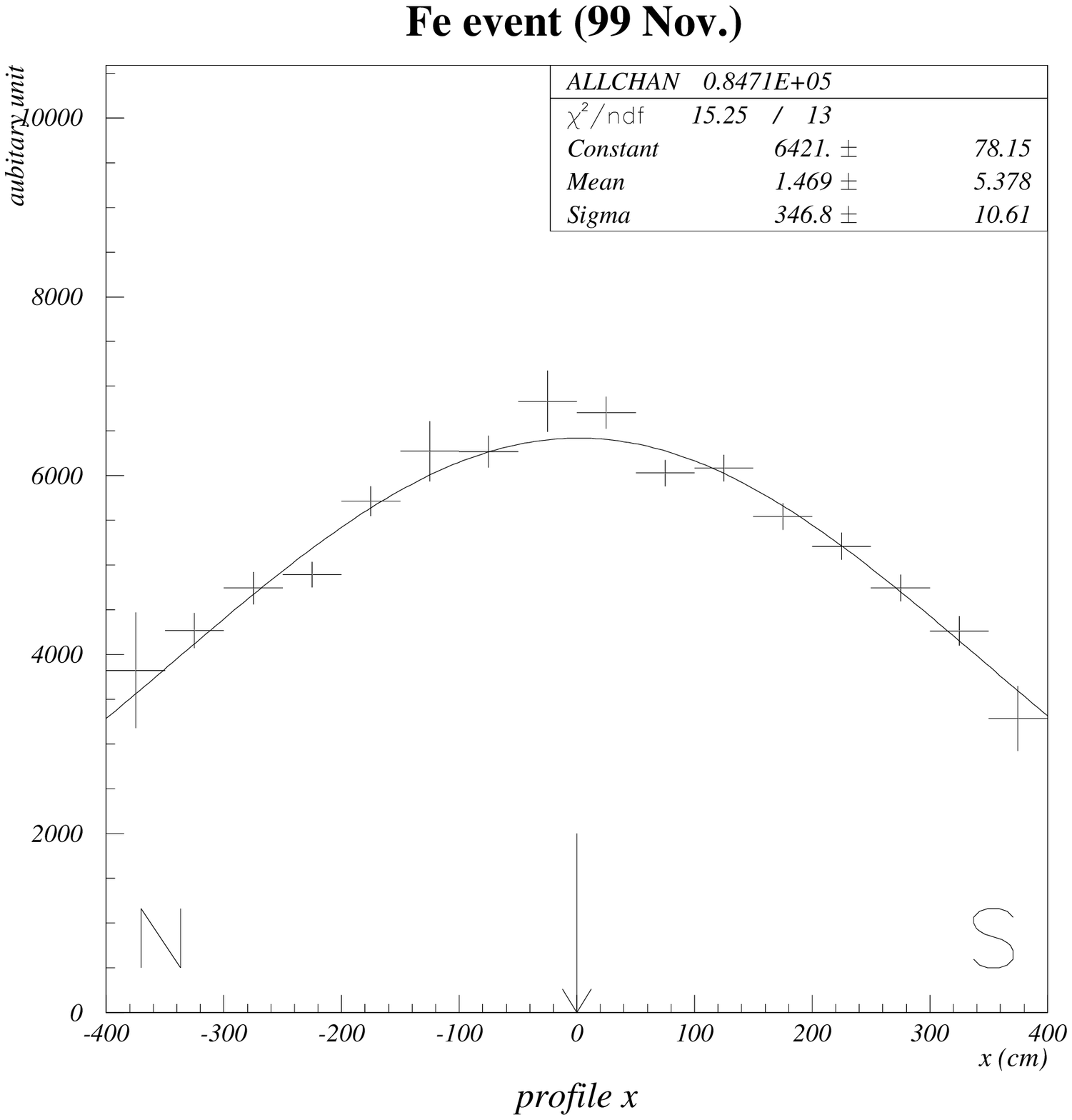,width=8cm}}
\centerline{\psfig{file=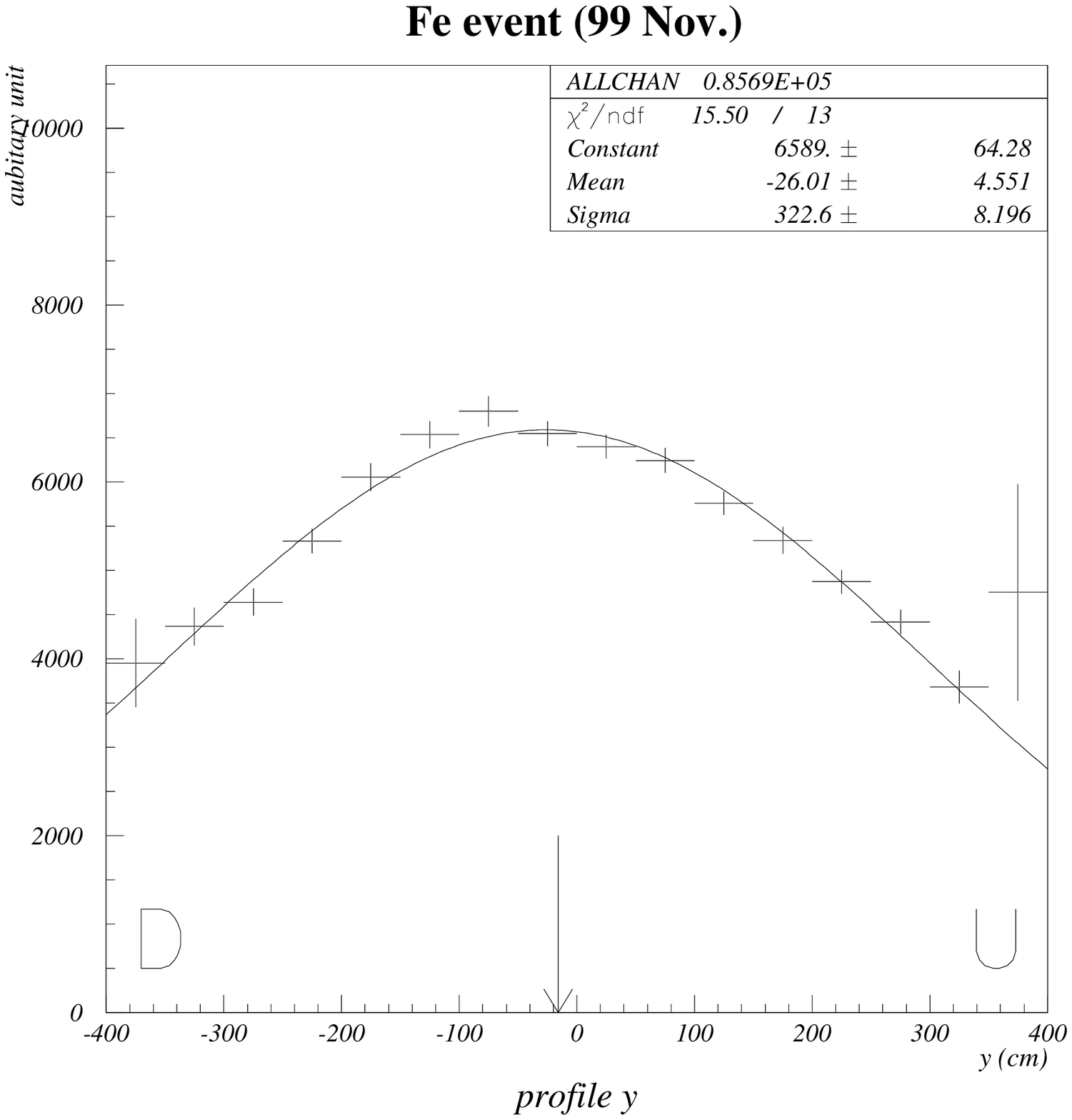,width=8cm}}
\caption{(top) the horizontal beam profile and (bottom) the vertical beam profile
in the muon ranger. The arrows indicate the direction to Super-Kamiokande}
\label{fig:muc1}
\end{figure}

\begin{figure}[p]
\centerline{\psfig{file=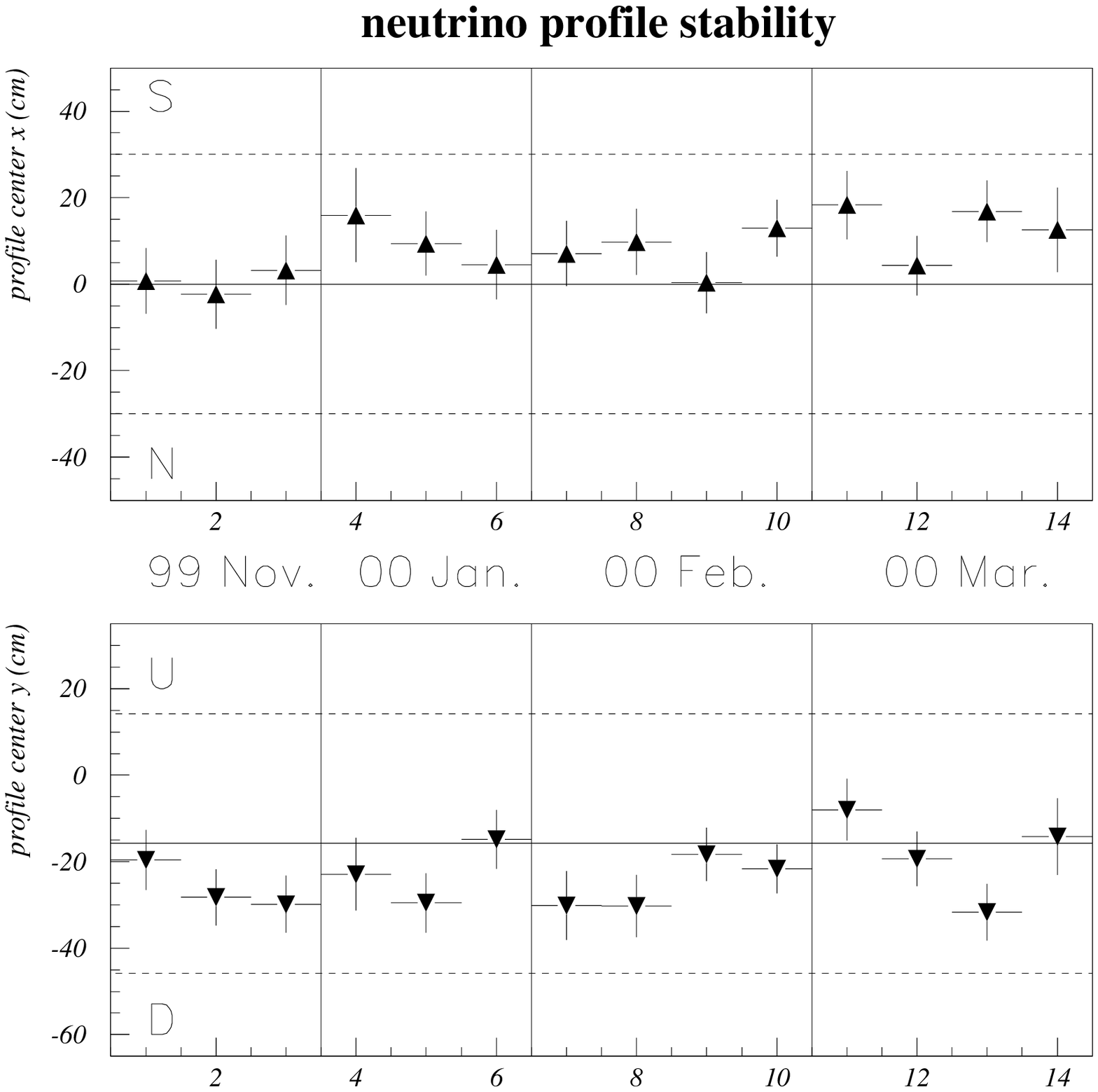,width=8cm}}
\caption{Stability of the center of the beam profile in the muon ranger over time. Each
point is represents the average beam profile center over a period of 2 days. The dashed
lines indicate the $\pm 1\;mrad$ deviation from the direction to Super-Kamiokande.}
\label{fig:muc2}\vskip0.5cm

\centerline{\psfig{file=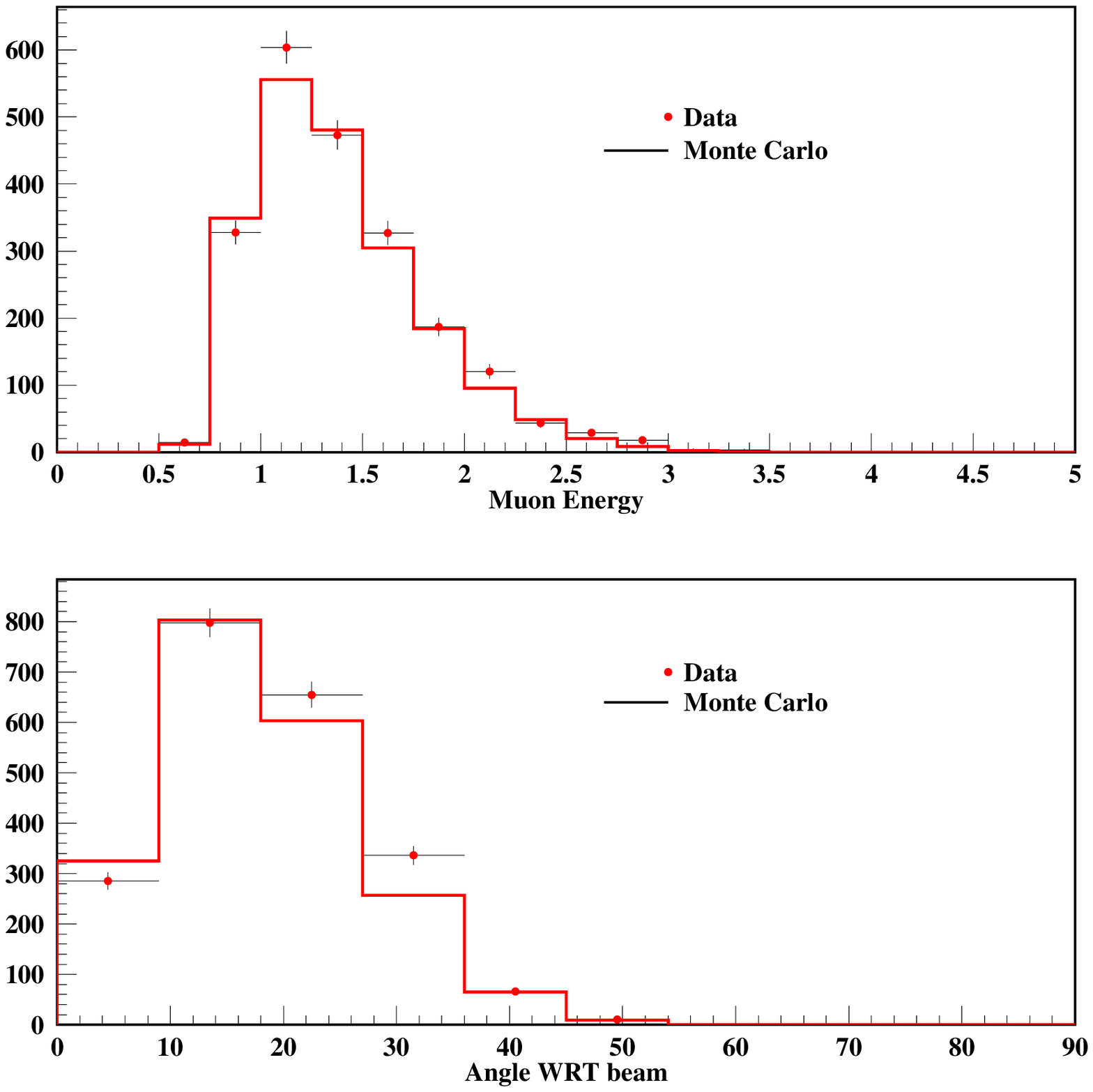,width=8cm}}
\caption{(a) Energy and (b) angular distribution of muons produced in neutrino
interactions in the Scifi compared to simulation.}
\label{fig:scifi1}
\end{figure}

 Comparison of the expected and observed number of events in Super-Kamiokande depends
on understanding the stability and characteristics of the neutrino beam profile and energy
spectrum.

The neutrino beam profile measured in the MRD is shown in Figure \ref{fig:muc1}.
The beam profile is centered on Super-Kamiokande and is reproduced well by simulation. The
Monte Carlo histogram has been normalised to the number of events in data.
The center of the beam profile is plotted as a function of time in Figure \ref{fig:muc2}.
The beam direction has been stable within $\pm1$ mrad throughout the running time of the experiment.
%The rate of neutrino interaction in the MRD is shown in Figure \ref{fig:muc3}.

 The energy and angular distributions of muons from neutrino interactions in the Scifi
is shown in Figure \ref{fig:scifi1}. The simulation has been normalised to the number
of events in data. The Monte Carlo shows good agreement with the data.

 The number of events predicted to be observed at Super-Kamiokande ($N_{SK}^{pred}$) in the absence of
neutrino oscillations is calculated using the observed number of events in the near detector 
($N_{KEK}^{obs}$)
corrected for the differences in neutrino flux, target mass, detector efficiencies and
livetimes between the near and far detector :
\begin{equation}
 N_{SK}^{pred} = N_{KEK}^{obs} \cdot R \cdot \frac{M_{SK}}{M_{KEK}} \cdot \frac{\epsilon_{SK}}{\epsilon_{KEK}}
\cdot \frac{L_{SK}}{L_{KEK}}
\end{equation}
where $\epsilon_{SK(KEK)}$ is the detection efficiency of neutrino interaction in 
Super-Kamiokande(KEK),
$M_{SK(KEK)}$ is the fiducial target mass at Super-Kamiokande(KEK), $L_{SK(KEK)}$ is
the number of protons on target collected by Super-Kamiokande(KEK) and 
R is the ratio of the number of events expected
at Super-Kamiokande and at KEK :
\begin{equation}
\begin{centering}
 R = \frac{\int \phi_{SK}(E_{\nu}) \sigma(E_{\nu}) dE}{\int \phi_{KEK}(E_{\nu}) \sigma(E_{\nu}) dE}
\end{centering}
\end{equation}
$R$ is calculated from the beam simulation.

\begin{table}
\caption{Number of events expected in Super-Kamiokande in the absence of oscillations
as estimated from each of the near detector components.}
\label{table:exp}
\begin{tabular}{lc}\hline\noalign{\smallskip}
    Detector    &    Expected number of events in SK \\ \hline\noalign{\smallskip}

      1kt       &       $40.3^{+4.7}_{-4.6}$         \\ \hline\noalign{\smallskip}
     Scifi      &       $40.1^{+4.9}_{-5.4}$         \\ \hline\noalign{\smallskip}
      MRD       &       $41.5^{+6.2}_{-6.4}$         \\ \hline\noalign{\smallskip}
\end{tabular}
\end{table}

The expected number of events observed in Super-Kamiokande 
based on the observed event rate in each of the near
detector components, corresponding to a collected intensity of $2.3 \cdot 10^{19}$ protons on target,
 are shown in Table \ref{table:exp}. The errors are systematic with
the statistical error being negligible.
Despite different detection techniques and sources of systematic error, the predictions of the
near detector components agree well. The error for the prediction using the 1kt detector is 
smaller than for the Scifi or MRD predictions,
since the 1kt detectors uses the same target matrial and the
same detection technique as used in Super-Kamiokande. Hence, the uncertainties in the neutrino-water 
interactions cross section and some detector-based biases cancel. For this reason, the prediction of the
1kt detector will be used when
comparing with the observed number of events in Super-Kamiokande. 
The largest sources of systematic error in the expected number of events derived from the 1kt event rate are
a 7\% error in $R$, the far-to-near ratio, and a 6\% error in the measurement
of the 1kt event rate. The extrapolation error arises from uncertainties in the pion monitor measurement
and in modelling low energy neutrino production. The error in the 1kt rate arises mostly from
uncertainties in vertex fitting and the propagation of these uncertainties into the fiducial volume
estimate.

\section{Events in Super-Kamiokande}
\label{sec:sk}

 KEK beam associated $\nu_{\mu}$ interactions in Super-Kamiokande are tagged by comparing
the UTC time stamps of the Super-Kamiokande trigger and the KEK beam spill. If the event
is indeed associated with the KEK beam, the difference between the two time stamps,
$\Delta T = T_{SK} - T_{KEK} - TOF$, where $TOF$ is the time the neutrino takes to move
between KEK and Super-Kamiokande, should be distributed between 0 and 1.1 $\mu s$, a 
reflection of the KEK beam spill width. The intrinsic time resolution of the GPS system
is approximately 150~ns so a 1.5 $\mu s$ search window is used. The
contamination from atmospheric neutrino events in the Super-Kamiokande sample 
is estimated to be less than $10^{-3}$ events over the
running period to date. 

\begin{figure}
\centerline{\psfig{file=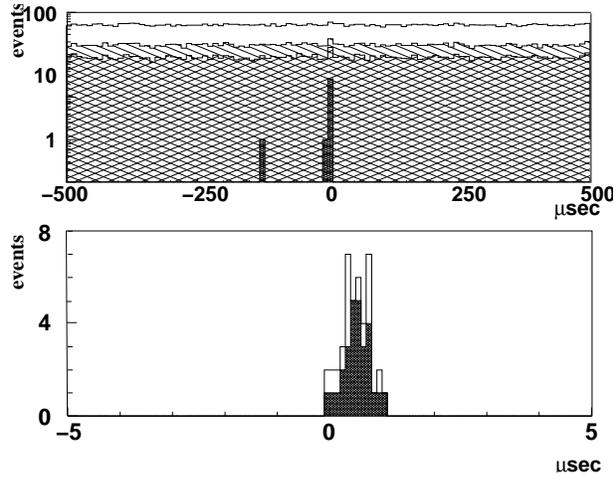,width=8cm}}
\caption{
(top) $\Delta(T)$ distribution for $\pm\!500\,\rm\mu{s}$ for all events
    (blank), after cuts 1) (hatched), 2) (double hatched) and
    3) (shaded) (see text).
    (bottom)
    Same for $\pm$5\,$\mu$sec for FC events only.
    The shaded area is for events inside the % 22.5\,kt
    fiducial volume.}
\label{fig:FCtime}
\end{figure}

 The events collected at Super-Kamiokande are sorted into two categories : fully-contained
(FC) events, in which the Cerenkov light from all particles in the event is contained in
the Super-Kamiokande inner detector, and outer-detector (OD) events, in which the interaction
occurs in the Super-Kamiokande anti-detector volume or in the rock surrounding the detector.
The FC events are furthur subdivided into FC-in(out) events, in which the interaction occurs
inside(outside) the 22.5 kt fiducial volume.
The reconstruction efficiency and systematics of the FC-in events are well known and it is
this category which will eventually be used in the full spectral analysis. The OD and FC-out 
events, which have larger
systematic errors, are used as consistency checks on the event rate
of the FC events. Other selection criteria for FC events are
\begin{enumerate}
\item no activity 30 $\mu$s before the event
\item $Q_{300ns}$, the total number of photo-electrons collected in the inner detector
      in a 300ns timing window, must be greater than 200 photoelectrons,
      corresponding to 30 MeV energy threshold.
\item $N_{OD}$, the number of hits in the largest cluster
      in the anti-detector must be less than 10.
\end{enumerate}
The selection efficiency of events in Super-Kamiokande is approximately 80\%, dominated by
the 30 MeV energy threshold.
Figure \ref{fig:FCtime} shows the FC event distribution within 500 $\mu s$ and 
5 $\mu s$ of the KEK beam spill.

\begin{table*}[t]
\caption{Summary of the observed number of events and the expected number of events
 at far site for the no oscillation hypothesis and three specific oscillation parameters. In
this table $sin^{2}(2\theta)=1$.}
\label{table:SK}
\begin{tabular}{lccccc}
\hline\noalign{\smallskip}
              &    Data  & No oscillation & $\Delta m^{2} = 3 \times 10^{-3} $ & $\Delta m^{2} = 5 \times 10^{-3}
$  & $\Delta m^{2} = 7 \times 10^{-3} $  \\
\hline\noalign{\smallskip}
1 ring $\mu$-like &  14    & $21.9\pm3.5$ & $12.4\pm2.1$ & $7.5\pm1.4$ & $6.8\pm1.2$ \\
1 ring $e$-like  &  1    & $2.5\pm0.5$ & $2.1\pm0.4$ & $1.9\pm0.4$ & $1.3\pm0.4$ \\
Multi Ring   &  12   & $16.0\pm2.7$ & $12.2\pm2.1$ & $8.4\pm1.5$ & $6.3\pm1.1$ \\
\hline\noalign{\smallskip}
FC-in & 27 & ${40.3}^{+4.7}_{-4.6}$ & $26.6^{+2.4}_{-2.2}$ & $17.8^{+2.3}_{-2.2}$ & $14.9^{+1.9}_{-1.9}$ \\
\hline\noalign{\smallskip}
\end{tabular}
\end{table*}

\begin{figure}[h]
\centerline{\psfig{file=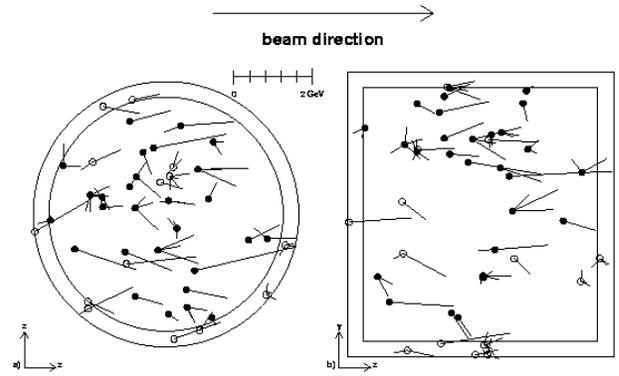,width=8cm}}\vskip1cm
\caption{The distribution of vertices of FC beam associated events in the Super-Kamiokande from the
top (a) and the side (b). The closed (open) circles indicate the position of the vertex in (out of)
the 22.5 kiloton fiducial volume (indicated by the inner circle). The lines show the fitted direction
of observed particles in the event. The direction of the KEK beam is horizontally left to right.}
\label{fig:vertices}
\end{figure}

A clear peak corresponding to beam-associated events may be observed from -0.2 to 1.3 $\mu s$.
The number of FC-in events and their characteristics are summarised in Table \ref{table:SK}. The
events are subdivided into events with a single Cerenkov ring (1-ring events) and multi-ring
events. The 1-ring events are furthur decomposed into ``mu-like'' and ``electron-like'' events. The
table also includes the expected number of events in each category for the case of no oscillations
and for three sets of oscillation parameters. Figure \ref{fig:vertices} shows the fitted vertex distribution
of all FC events in Super-Kamiokande. The direction and length of the lines on each vertex
indicate the direction and momentum of each detected particle in the event. There is an obvious momentum
flow in the direction of the KEK beam.

\section{Summary}
\label{sec:concl}

The K2K experiment has been running smoothly since June, 1999, and
collected  $2.6 \times 10^{19}$ protons on target.
The expected interaction rates at Super-Kamiokande as determined by the near detectors (SciFi, MRD and
1kton) are consistent with one another.
The neutrino beam direction has been within $1\,mrad$ of the direction to Super-Kamiokande throughout
the experiment.
Twenty seven KEK beam associated FC-in events have been observed in the
Super-Kamiokande fiducial volume with a background of less than $10^{-3}$ events. The expected
number is 
$40.3^{+4.7}_{-4.6}$ in the case of null oscillations.
Our data disfavor null oscillations 
at the $2\sigma$ level. 
With more data a definitive analysis of both the event rate and reconstructed energy spectrum
of the events in the far detector will be carried out.

\end{document}